\documentclass[a4paper,nofootinbib,preprintnumbers,floatfix,superscriptaddress,prl,twocolumn,showpacs]{revtex4-1}
\usepackage{amsmath, amsthm, amssymb}
\usepackage{graphicx}
\usepackage{dcolumn}
\usepackage{bm}
\usepackage{color}
\usepackage{amsmath}
\usepackage{amsfonts}
\usepackage{amssymb}
\usepackage{bbm}
\usepackage{hyperref}

\newcommand{\eqnref}[1]{(\ref{#1})}
\newcommand{\figref}[1]{Fig.~\ref{#1}}

\newcommand{\vacr}{|vac\rangle}
\newcommand{\vacl}{\langle vac|}
\newcommand{\bra}[1]{\langle #1|}
\newcommand{\ket}[1]{|#1\rangle}

\DeclareMathOperator{\Tr}{Tr}

{\begin{framed}\begin{small}}
{\end{small}\end{framed}}

\graphicspath{{./Figs/}}
\DeclareGraphicsExtensions{.pdf,.png,.jpg}

\makeindex

\begin{document}

\title{Robust nonlocality tests with displacement-based measurements}
\date{\today}

\author{Jonatan Bohr Brask}\affiliation{ICFO-Institut de Ciencies Fotoniques, Av.~Carl Friedrich Gauss, 3, 08860 Castelldefels (Barcelona), Spain}
\author{Rafael Chaves}\affiliation{ICFO-Institut de Ciencies Fotoniques, Av.~Carl Friedrich Gauss, 3, 08860 Castelldefels (Barcelona), Spain}

\begin{abstract}
Lately, much interest has been directed towards designing setups that achieve decisive tests of local realism. Here we present Bell tests with measurements based on linear optical displacements and single-photon detection. The scheme displays good tolerance to loss. In particular, for entangled squeezed states, we find thresholds compatible with current efficiencies of detectors and sources. Furthermore, the scheme is easily extendible to any number of observers, allowing observation of multipartite nonlocality for a single photon shared among multiple modes. We also consider the case of atom-photon entanglement, where the loss threshold can be lowered further, as well as local filters compensating transmission and coupling inefficiencies at the source.
\end{abstract}

%\pacs{}

\maketitle

\emph{Introduction.---}%
Experiments performed by space-like separated, independent observers may display correlations that do not comply with assumptions of local realism, i.e.~that physical quantities have well established values prior to measurement and that signal propagation is restricted by the speed of light in accordance with special relativity. Such assumptions lead to restrictions, usually expressed as Bell inequalities~\cite{bell1964}, which can be surpassed within quantum theory. Correlations which violate a Bell inequality are referred to as nonlocal. From an applied point of view, nonlocality represents a physical resource, which enables protocols such as device-independent quantum key distribution~\cite{QKD} and random number generation~\cite{random}. These protocols rely on the violation of a Bell inequality to ensure secrecy and randomness, without further assumptions about their physical implementation.

In spite of steady theoretical and experimental progress, nonlocality has yet to be demonstrated in a loophole-free manner. All experiments to date suffer from either the locality loophole, meaning that the assumption of space-like separation of the observers is not fulfilled~\cite{rowe2001,*matsukevich2008}, or the detection loophole, which is opened when the efficiency of the detectors employed in the Bell test is insufficient~\cite{aspect1982}. Both loopholes admit local-hidden-variable explanations of the observed data and compromise the security of cryptographic protocols. Closing the loopholes is important both from a fundamental and a technological perspective. To close the locality loophole, it is advantageous to work with optical systems since light can be distributed with relative ease among spatially separated parties and since optical detectors are fast. Various approaches have been considered towards closing the detection loophole in optical Bell tests. Two fundamental types of entangled states which may be used are polarisation-entangled states of fixed photon number, or states relying on superposition of one or few photons with the vacuum. Both approaches are hampered by the low efficiency of most available single-photon detectors  \cite{hadfield2009} (although high-efficiency superconducting transition-edge sensors are becoming more widespread). The former case has the advantage that projective measurements in any basis can be performed with linear optics. However, generating entangled states of more than two modes efficiently is highly challenging. For the case of photon-number superpositions, entanglement generation can be achieved with relatively large efficiency since, in the simplest cases, it suffices to produce a two-mode squeezed state, or to split a single photon using beam splitters. The disadvantage is that perfect projective measurements are not available in all bases, using linear optics and photon counting, since passive linear optics cannot change the energy of a state.

Here we demonstrate that using the simple, displacement-based measurements of Ref.~\cite{banaszek1999}, it is possible to attain good efficiency thresholds which in some cases exactly coincide with the thresholds for perfect measurements in arbitrary bases. In an all-optical implementation, our scheme works for a two-mode squeezed state and admits a combined efficiency threshold for coupling, transmission, and detection of $66.7\%$. The scheme can be extended to detect multipartite nonlocality for $N$ parties by applying it to a single photon split between multiple modes, i.e.~a single-photon W-state \cite{Wstate}, possibly entangled with an atom. The W-state is known to be extremely robust against losses \cite{Wrobust}, and we demonstrate efficiency thresholds of $\sim 68\%$ for the purely optical state in the limit of high $N$ and of $50\%$ for the atom-photon case independent of $N$. For bipartite atom-photon entanglement, the threshold can be lowered to $43.7\%$ using additional measurement settings. Furthermore, we suggest to compensate transmission and coupling losses at the source by local filtering. Surprisingly, the use of filters allows Bell inequality violations for arbitrarily low atom-photon coupling efficiency.

\begin{figure}[t!]
\includegraphics[width=.485\textwidth]{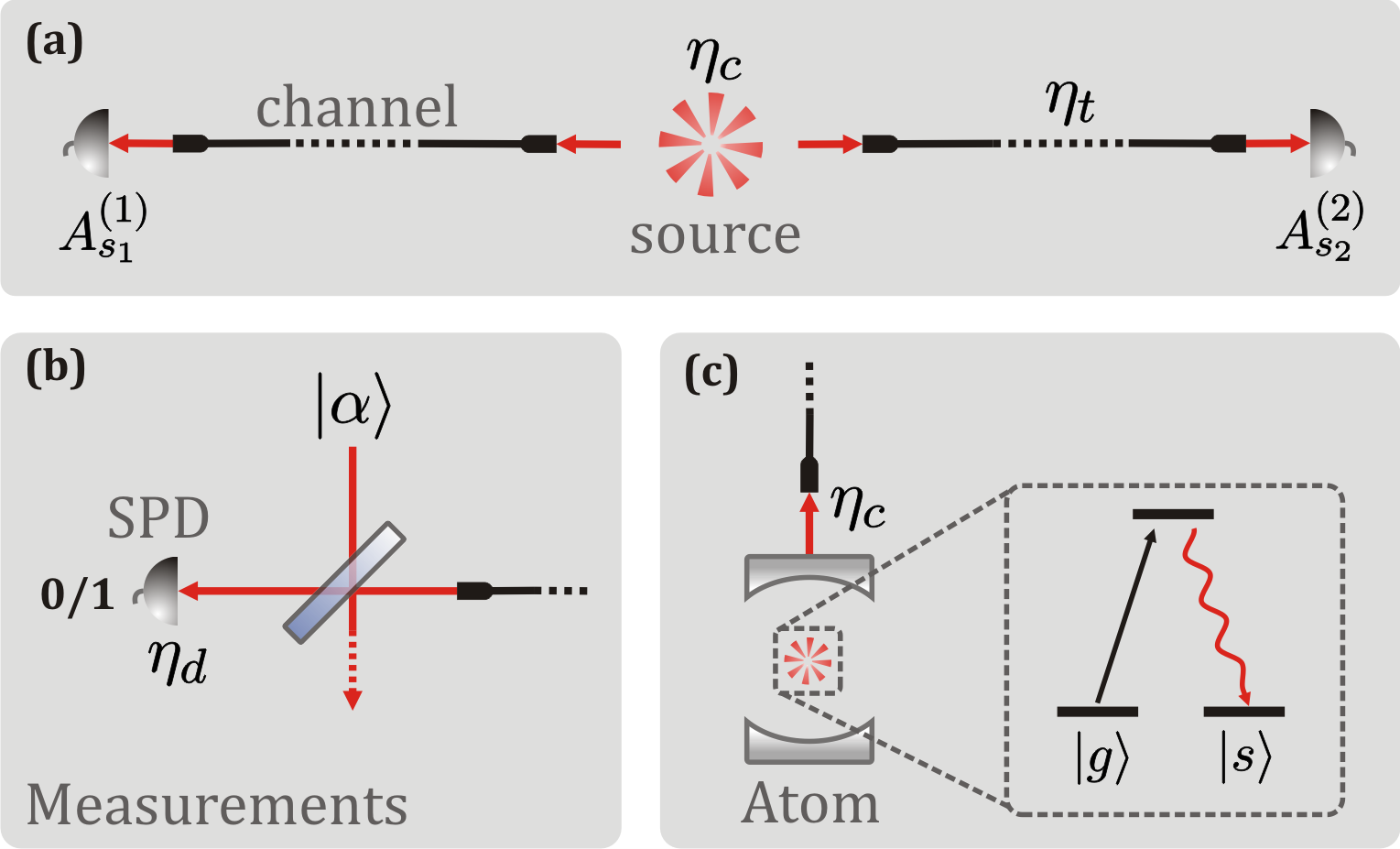}
\caption{(Colour online) \textbf{(a)} A source with coupling efficiency $\eta_c$ distributes an entangled state to two parties through lossy channels of transmissivity $\eta_t$. Each party chooses between two measurements with binary outcomes. \textbf{(b)} Measurements consists in a variable displacement, followed by single-photon detection by bucket detectors of efficiency $\eta_d$. The displacement controls the measurement setting. \textbf{(c)} Atom-photon entanglement. We consider an atom with a $\lambda$-type level scheme. The atom is initially prepared in one of two stable ground states $\ket{g}$. An entangled state between atom and light is then generated by partly exciting the atom allowing decay into $\ket{s}$ accompanied by emission of a photon. The photon is coupled out of the cavity and distributed to the remaining party.} \label{fig.setup}
\end{figure}

\emph{Setup and measurements.---}%
Throughout most of the paper, we will consider a bipartite scenario with each party $k$ having a choice of $m$ dichotomic measurements $A^{(k)}_0$, $A^{(k)}_1$, as illustrated in \figref{fig.setup}(a). For $m=2$, the only relevant Bell inequality is the CHSH inequality \cite{CHSH}
\begin{equation}
\label{eq.chsh}
\langle A^{(1)}_0 A^{(2)}_0 \rangle + \langle A^{(1)}_0 A^{(2)}_1 \rangle + \langle A^{(1)}_1 A^{(2)}_0 \rangle - \langle A^{(1)}_1 A^{(2)}_1 \rangle \leq 2 .
\end{equation}
In the case of $m \geq 3$, we consider the I$_{mm22}$ family of inequalities \cite{collins2004}.

As shown in \figref{fig.setup}(a), a source produces an entangled state, which is distributed to the parties through lossy channels, e.g.~optical fibres, of transmittivity $\eta_t$. Each party performs a measurement of the type suggested in Ref.~\cite{banaszek1999}, namely a displacement followed by single-photon detection. Physically, the displacements are implemented by mixing the signal with a coherent state from a local oscillator on a beam splitter with very high transmission, as shown in \figref{fig.setup}(b). We take the single-photon detectors to have an efficiency $\eta_d$, and they do not need to be number resolving. Different measurement settings correspond to different choices for the displacement, i.e.~different phase and intensity of the local oscillator. A lossy detector can be modeled as a perfect detector preceded by loss acting on the input signal. This loss can be commuted through the displacement, since any loss from the local oscillator can be compensated by increasing the amplitude and is thus irrelevant. Hence, as the detector loss is the same for all settings, it is equivalent to transmission loss and we can account for it by modifying the state, leaving the detection ideal. For a measurement by a perfect single-photon detector preceded by a displacement $D(\alpha)$, the no-click outcome corresponds to the projector $P_0 = D(\alpha)\ket{0}\bra{0}D^\dagger(\alpha) = \ket{\alpha}\bra{\alpha}$, where $\ket{\alpha}$ denotes a coherent state. Assigning $+1$/$-1$ to the no-click and click events respectively, the measurement operator is then given by $M(\alpha) = P_0 - (1-P_0) = 2\ket{\alpha}\bra{\alpha} - 1$, with matrix elements in the Fock basis
\begin{equation}
\label{eq.dispmeasop}
M_{nn'}(\alpha) = 2 e^{-|\alpha|^2} \frac{\alpha^{n'}(\alpha^*)^n}{\sqrt{n!n'!}} - \delta_{nn'} .
\end{equation}

\emph{Bipartite states and thresholds.---}%
We first apply our scheme to a two-mode squeezed vacuum state. This is perhaps the simplest entangled state of light to produce, and in the Fock basis it takes the form
\begin{equation}
\label{eq.tmss}
\sqrt{1-\lambda^2} \sum_n \lambda^n e^{i\phi n} \ket{n,n} ,
\end{equation}
where $\lambda$ sets the magnitude and $\phi$ the direction of squeezing. The state \eqnref{eq.tmss} can be produced e.g.~by non-degenerate parametric down conversion in a nonlinear crystal, or by mixing two single-mode squeezed states on a balanced beam splitter.
When subjected to loss, the state becomes \cite{filip2002}
\begin{align}
\label{eq.lossytmss}
& \rho_{sq}(\eta) = (1 \! - \! \lambda^2) \sum_{n,n'} (\lambda \, \eta)^{n+n'} e^{i\phi (n' \! - \! n)} \! \sum_{k,k'}^{\min n,n'} \left(\frac{1 \! - \! \eta}{\eta}\right)^{k+k'} \nonumber \\
& \times \! \sqrt{ {n \choose k}\!{n \choose k'}\!{n' \choose k}\!{n' \choose k'} }  \ket{n \! - \! k,n \! - \! k'}\bra{n' \! - \! k,n' \! - \! k'} . \!
\end{align}
As mentioned above, detector efficiency can be included in the transmission efficiency in our scheme. The same is true for coupling efficiency $\eta_c$ at the source. Thus we can account for all losses by taking $\eta = \eta_c\eta_t\eta_d$. Using \eqnref{eq.lossytmss} and \eqnref{eq.dispmeasop}, we can compute all the correlators $\langle A^{(1)}_{s_1} A^{(2)}_{s_2} \rangle = \Tr \left[ M(\alpha_{s_1}) \otimes M(\alpha_{s_2}) \rho_{sq}(\eta) \right]$ appearing in \eqnref{eq.chsh}. By numerical maximisation over the settings, we find that CHSH can be violated above the threshold $\tilde{\eta} = 66.7\%$. Surprisingly, this coincides exactly with the lowest possible threshold attainable for \textit{any} state and \textit{any} measurements in the CHSH scenario, when the losses for both parties and measurement settings are equal \cite{wilms2008}. Thus no state preparation and no measurements, no matter how complicated, can achieve a better loss tolerance than our simple setup. The threshold can be reached taking the same settings for both parties, with the $\alpha$'s real, and $\phi = \pi$. The required squeezing and displacements are small, vanishing at the threshold. For e.g.~$\eta = 67\%$ we need a squeezing of $\sim 0.1$dB and average photon numbers $|\alpha_0|^2 \sim 10^{-2}$ and $|\alpha_1|^2 \sim 10^{-6}$. We have checked that the scheme is robust against fluctuations in the $\alpha$'s, with a 10\% deviation from the optimal values leading to a 1\% increase of $\tilde{\eta}$. We remark that we have also allowed for additional measurement settings by testing the I$_{mm22}$ family of inequalities with $m$ settings per party \cite{collins2004}. However, we obtained no further improvement of $\tilde{\eta}$ for $m \leq 5$.

Note that ours is not the first scheme to reach the optimal threshold of $66.7\%$. The threshold can be reached with polarisation-entangled states and polarisation measurements, as shown by Eberhard \cite{eberhard1993}. The states in the Eberhard scheme can be implemented using a squeezing source similar to ours, with some additional manipulation to control the degree of polarisation entanglement, see e.g. \cite{thomas2010}. However, the present scheme demonstrates that the optimal threshold can be reached also in the absence of Pauli-type measurements and provides a simple alternative setup. Our threshold represents a significant improvement over the recent schemes of Refs.~\cite{cavalcanti2011,chaves2011}. In particular, in~\cite{cavalcanti2011} transmission and detector efficiencies are not equivalent. Even assuming perfect homodyne detection, and perfect coupling and transmission, the critical single-photon detector efficiency is $71.1\%$. For a finite coupling and transmission efficiency $\eta_c\eta_t=90\%$, still assuming perfect homodyning, the threshold in the current scheme is lower by 15\%. Two-mode squeezed states have been generated with efficiency 85\% \cite{eisaman2011,fedrizzi2007} and recent experiments suggest that efficiencies exceeding 90\% may be possible \cite{takahashi2008,gerrits2011}. With transmission losses of a few percent the required detector efficiency is then on the order of 75\%. This number is reachable by commercial detectors, and is well below the efficiency of superconducting transition-edge sensors. Hence, an implementation of our setup seems within reach of current technology.

Next, we consider replacing one of the parties by an atom in a cavity on which projective measurements in any basis can be performed with very high accuracy~\cite{rowe2001,*matsukevich2008}. Such a setup has been studied in Refs.~\cite{sangouard2011,chaves2011} and is shown in \figref{fig.setup}(c). Taking into account a finite coupling efficiency for the emitted photon into the desired mode, and finite transmission and single-photon detection efficiencies, the generated joint state of atom and light is of the form $\rho_{at}(\eta_c\eta_t\eta_d)$, with
\begin{align}
\rho_{at}(\eta) = & [ \cos(\theta)\ket{g}\ket{0} + \sqrt{\eta} \sin(\theta)\ket{s}\ket{1} ]\otimes h.c \nonumber \\
 & + (1-\eta) \ket{s}\vacr \otimes h.c. ,
\end{align}
where the coefficient $\cos(\theta)$ is determined by the strength of the initial excitation pulse and $h.c.$ denotes hermitian conjugate. For the optical mode, the protocol proceeds exactly as before, with measurements of the type in \figref{fig.setup}(b). For the atomic party we allow projective measurements along any arbitrary direction on the Bloch sphere, and we take the detection efficiency to be unity. For the CHSH inequality, the best possible threshold in this case is $\tilde{\eta} = 50\%$ \cite{wilms2008}. Remarkably, it is possible to reach this bound exactly using displacement-based measurements. However, we can improve further by allowing a third measurement setting $A^{(k)}_2$ for both parties. We have tested the I$_{3322}$ inequality \cite{collins2004}, which applies to this scenario, and we find $\tilde{\eta} = 43.7\%$. An additional measurement setting does not complicate the experiment. Thus, with a trapped atom, this is the most attractive setup to implement. We remark that again the threshold obtained with the displacement scheme is very close to the bound obtained with perfect Pauli measurements, which is 43\% \cite{brunner2007}. Adding more settings and testing I$_{mm22}$ does not appear to improve the threshold.

\emph{Multiple parties and filtering.---}%
We now consider two possible modifications of our scheme. Extension to a multipartite nonlocality test, and local filtering to improve the tolerance to losses.

For a scenario with $N$ parties, each having a choice of two dichotomic measurements, there are $2^{2^N}$ tight, linear, full-correlator Bell inequalities, all of which can be compactly expressed in the single, nonlinear Werner-Wolf-Weinfurter-Zukowski-Brukner (W${}^3$ZB) inequality \cite{werner2001,*weinfurter2001,*zukowski2002}
\begin{equation}
\label{eq.WW}
\sum_{r} \left\vert \tilde{\xi}(\bf{r}) \right\vert \leq 1 ,
\end{equation}
where $\tilde{\xi}({\bf r})=2^{-N}\sum_{{\bf s}} (-1)^{{\bf r}\cdot{\bf s}} \xi({\bf s})$, with ${\bf r}$ and ${\bf s}$ being vectors in $\{0,1\}^N$ and $\xi({\bf s}) =\xi(s_1,\ldots,s_N)= \langle A^{(1)}_{s_1} \cdots A^{(N)}_{s_N} \rangle$ the corresponding full correlator. We have tested this inequality using the measurements of \figref{fig.setup}(b). A simple way to create an entangled state of $N$ parties is by splitting a single photon between $N$ modes using suitable adjusted beam splitters. A single photon can be generated by conditioning on the detection of a photon at one output of a two-mode squeezing source. The result is a W-state
\begin{equation}
\ket{W_N} = \frac{1}{\sqrt{N}}\left( \ket{1,0,\ldots,0} + \ldots + \ket{0,0,\ldots,1} \right).
\end{equation}
When undergoing loss in each mode, this state is transformed into $\rho_{W_N}(\eta) = \eta \ket{W_N}\bra{W_N} + (1-\eta) \vacr\vacl$, where $\vacr$ denotes a state with no photons. For perfect Pauli measurements, $\rho_{W_N}(\eta)$ violates \eqnref{eq.WW} for any number of parties, and the critical $\eta$ required for violation improves with $N$ as $\tilde{\eta} = 2/(3-N^{-1})$, giving an asymptotic threshold $\tilde{\eta} \to 66.7\%$ for $N\rightarrow\infty$. For qubits encoded into zero- and single-photon Fock states, perfect projections generally are not available. However, good tolerance to losses can be obtained with displacement-based measurements.

Using \eqnref{eq.dispmeasop}, for the simplest case where all parties use the same settings, we analytically compute the full correlator $\xi({\bf s})$ in the state $\rho_{W_N}(\eta)$. We then numerically obtain the critical efficiency $\tilde{\eta}$ for violation of W${}^3$ZB. We have verified for $N\leq 10$, that allowing different settings for each party does not improve $\tilde{\eta}$. We find that the qualitative behaviour of $\tilde{\eta}$ is similar to the ideal case of perfect Pauli measurements. For small $N$ it decreases rapidly, then approaches an asymptotic value. Unfortunately, we have not been able to obtain an analytical expression for the threshold, however, we have fitted a function of the form $a/(b-N^{-1})$ to our data. The critical efficiency is well described by this with $a=1.841$ and $b=2.696$, which leads to an asymptotic value of $\tilde{\eta} \to 68.3\%$. The rapid decrease for small $N$ is nice from an experimental perspective, since it means that investigation of multipartite nonlocality becomes possible with relatively few measurement stations. E.g., for $N=5$ we already have $\tilde{\eta} = 73.9\%$. When one party is replaced by an atom in a cavity, as considered above, the threshold becomes independent of $N$. We find $\tilde{\eta} = 50\%$ for any number of parties, which coincides with the bound for perfect Pauli projectors. The present scheme again represent a significant improvement over previous proposals. In Ref.~\cite{chaves2011}, the required detection efficiency necessary to violate W${}^3$ZB with an optical W-state is $\sim 85\%$. In proposals based on GHZ states \cite{cabello2008, he2009}, loss and detection-efficiency thresholds can compete with the ones obtained here. However, GHZ states are significantly more challenging to generate experimentally than W-states and can so far be produced only with small efficiencies, far from the required thresholds for loophole free Bell tests.

The second extension of our scheme aims to eliminate transmission and source-coupling losses by the use of local filtering. A similar idea was the basis for Ref.~\cite{gisin2010}, where it was used to improve Bell inequality violation as the basis for quantum key distribution. We apply this idea to the case of atom-photon entanglement and to the multipartite W-state. Since the effect of loss is to decrease the single-photon component of the state, we let each party apply the probabilistic single-photon amplifier, proposed by Ralph and Lund \cite{ralph2009} (an implementation with corresponding Kraus representation is llustrated in the Appendix). Since the amplification is probabilistic, it may sometimes fail. The parties only proceed with the protocol upon successful filtering. That is, they only proceed to choose the measurement basis after the filter has succeeded. Thus the filter can be seen as part of the state preparation. We can model the filter, acting on each mode of the state received by the parties, as a quantum channel. Taking the limit $t \to 0$ and renormalising, we find that a successful application of the filter by every party takes $\rho_{W_N}(\eta_c\eta_t\eta_d) \to \rho_{W_N}(\eta_c'\eta_d)$. That is, successful filtering compensates transmission and coupling loss between the source and the parties but introduces an additional loss, corresponding to the coupling inside the filter. Thus, as long as $\eta_c' < \eta_c\eta_t$, it is beneficial to apply the filters. For example, if the same type of single-photon generation is employed in the initial source and the filter (e.g. parametric down-conversion), $\eta_c'=\eta_c$ and it will be beneficial to filter as soon as there is significant transmission loss. However, successful preselection requires successful filtering by all parties simultaneously, and hence for a given success probability $p_f$, the rate of the experiment will decrease as $p_f^N$. Longer data collection times become necessary as the number of parties increases, and detector dark counts may become problematic.

In the case of atom-photon entanglement, the filtering is applied only to the photonic modes. For the bipartite setting, we find that by choosing an appropriate small value of $\theta$, for efficiencies close to the threshold $\tilde{\eta} = 43.7\%$, it is indeed possible to replace $\rho_{at}(\eta_c\eta_t\eta_d) \to \rho_{at}(\eta_c'\eta_d)$. This is a nice result. It says that, by applying local filtering, it is possible to break I$_{3322}$ for \textit{any} value of the atom-photon coupling efficiency, as long as the combined detection and single-photon production efficiency employed by the photonic party fulfills $\eta_c'\eta_d > 43.7\%$. For a bipartite test, a small value of the preselection probability $p_f$ is not critical, and a probabilistic single-photon source can be used. Efficient probabilistic single-photon generation is likely to be much easier to achieve in experiment than a high collection efficiency for atomic emission. For trapped ion or atoms, a collection efficiency of $\eta_c=50\%$ is an optimistic value at current (a fraction of spontaneous emission into the cavity mode of $51\%$ has been achieved \cite{mundt2001}, and a recent study predicts $\eta_c > 30\%$ for coupling into single-mode fibre \cite{kim2011}). On the other hand, single-photon sources based on parametric down conversion achieving $\eta_c' \sim 85\%$ have been demonstrated \cite{eisaman2011}. It thus becomes possible to perform the Bell test with detectors of very moderate efficiencies $\eta_d = 0.437/\eta_c' \sim 51\%$. In a recent experiment, readout from a single trapped atom with efficiency exceeding 98\% was demonstrated \cite{henkel2010}. We remark that detection efficiencies up to $99.99\%$ have been demonstrated for trapped ions \cite{myerson2008}.

\emph{Conclusion.---}%
In summary, we have presented a scheme for loophole-free violation of local realism based on simple measurements consisting of displacements followed by single-photon detection which does not need to resolve the photon number. We find favourable thresholds for the combined coupling, transmission, and detection losses. For a two-mode squeezed state, we reach $\tilde{\eta} = 66.7\%$, which is the lowest possible threshold for arbitrary states and measurements in the scenario with two parties, two dichotomich measurements and symmetric losses \cite{wilms2008}. For atom-photon entanglement we have shown that a threshold of $\tilde{\eta} = 47.3\%$ is achievable, when the atomic detection efficiency is high. We have extended our scheme to arbitrary party number $N$. In this case we find an asymptotic threshold of $\tilde{\eta} \sim 68\%$ for an all-optical W-state with large $N$ and $\tilde{\eta} = 50\%$ independent of $N$, when one party is replaced by a trapped atom. Local filtering can compensate losses at the source, and in particular we have seen that the use of local filters based on single-photon sources allows violations for an arbitrary small atom-photon coupling. The present scheme is implementable by currently available techniques, and thus we believe this work paves a feasible way towards experimental violation of local realism closing, simultaneously, both the detection and locality loopholes.

\emph{Acknowledgements.---}%
We thank C.~I.~Osorio, J.~Neergaard-Nielsen, and A. Ac\'in for fruitful discussions. J.~B.~Brask was supported by the Carlsberg Foundation and ERC starting grant PERCENT. R.~Chaves was supported by the Q-Essence project.

\begin{figure}
\includegraphics[width=.485\textwidth]{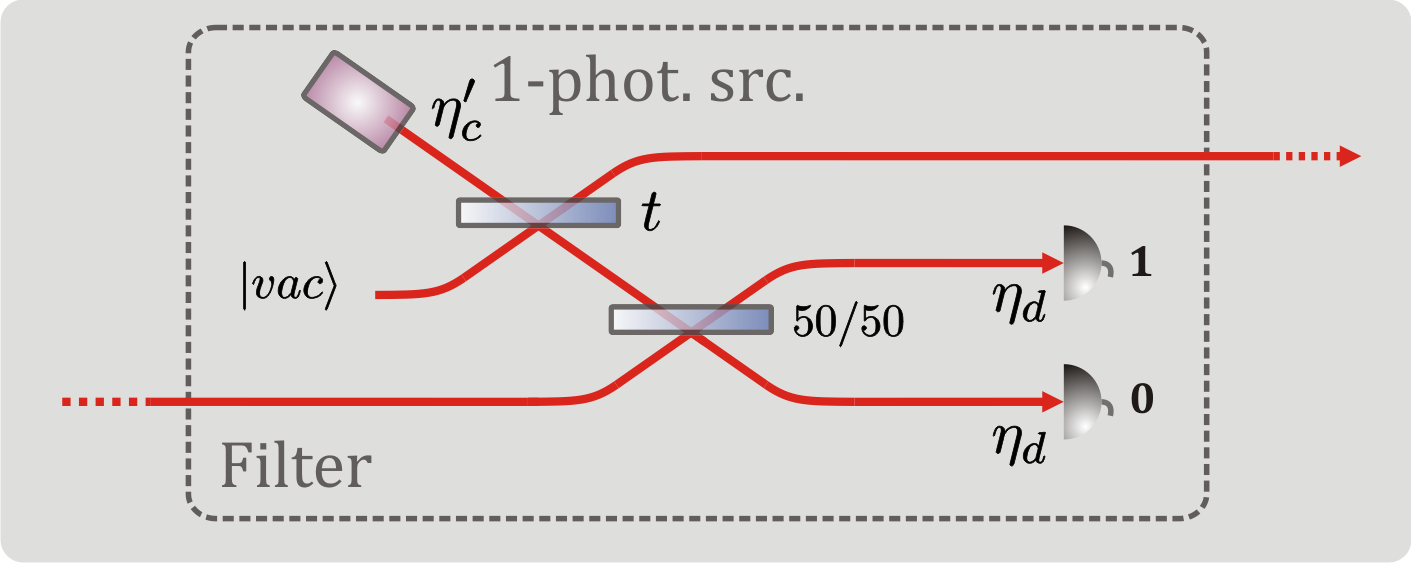}
\caption{(Colour online) Single-photon amplifier as a filter. Two auxiliary modes are used, one of which must contain a single photon. Success is conditioned on a click in one detector only. When the transmissivity $t$ is small, the single-photon component of the output is amplified, compensating preceding losses.} \label{fig.filter}
\end{figure}

\emph{Appendix.---}%
An implementation of the filter based on single photon amplification is shown in \figref{fig.filter}. Augmenting the circuit in the figure with additional vacuum modes and modeling losses by beam splitters, we can derive Kraus operators \cite{NandC}. We find that in the 0-1-photon subspace, up to normalisation, the channel is described by the four operators
\begin{align}
K_1 & = \sqrt{\frac{\eta_d (1-\eta_c'\eta_d)}{2}} \ket{0}\bra{1} ,  \\ \nonumber
K_2 & = - \sqrt{\frac{(1-t)\eta_c'\eta_d^2}{2}} \ket{0}\bra{1} , \\ \nonumber
K_3 & = - \sqrt{\frac{(1-t)\eta_c'\eta_d(1-\eta_d)}{2}} \ket{0}\bra{1} , \\  \nonumber
K_4 & = - \sqrt{\frac{(1-t) \eta_c'\eta_d}{2}} \ket{0}\bra{0} +
\sqrt{\frac{t \eta_c'\eta_d^2}{2}} \ket{1}\bra{1} .
\end{align}

%\bibliography{displacementscheme}


\end{document}